\documentclass[reqno,11pt]{amsart}
\newtheorem{prop}{Proposition}
\newtheorem{cor}{Corollary}
\newtheorem{df}{Definition}

\newcommand\eps\varepsilon
\newcommand\ph\varphi
\newcommand\kap\varkappa

\begin{document}

\title[Lagrangian Systems with non-smooth constraints]
{Lagrangian Systems with non-smooth constraints}

\author[Andrey Volkov, Oleg Zubelevich]{ Andrey Volkov\\ {\\\tt Dept. of Theoretical mechanics,  \\ Machine-building technologies and  equipment Faculty\\ Moscow State Technological University "Stankin"\\Russia, 127994, Moscow, Vadkovskii Lane, 1}\\\\ Oleg Zubelevich\\ {\\\tt
 Dept. of Theoretical mechanics,  \\
Mechanics and Mathematics Faculty,\\
M. V. Lomonosov moscow State University\\
Russia, 119899, moscow, Vorob'evy gory, MGU} \\
 }
%\email{ozubel@yandex.ru}
%\curraddr{2-nd  Krestovskii Pereulok 12-179, 129110, Moscow, Russia}
\date{}
%\thanks{Partially supported by grants
% RFBR 08-01-00681,  Science Sch.-8784.2010.1}

\begin{abstract}The Lagrange-d'Alembert equations with constraints belonging to $H^{1,\infty}$ have been considered. A concept of weak solutions to these equations has been  built. A global existence  theorem for Cauchy  problem   has been obtained. 
\end{abstract}

\maketitle
\numberwithin{equation}{section}
\newtheorem{theorem}{Theorem}[section]
\newtheorem{lemma}[theorem]{Lemma}
\newtheorem{definition}{Definition}[section]

\section{Introduction. The Statement of The Problem}
In this article we consider a dynamical  system with configuration space $M\subset\mathbb{R}^m=\{x=(x^1,\ldots,x^m)^T\}$ being  a bounded domain. 

The system is described by the Lagrangian
$$
L(x,\dot x)=T(x,\dot x)-V(x),\quad T=\frac{1}{2}g_{ij}(x)\dot x^i\dot x^j=\frac{1}{2}\dot x^TG(x)\dot x,\quad x\in M.$$
and (possibly non-holonomic) constraints
\begin{equation}\label{vt55v} a(x)\dot x=a^k_l(x)\dot x^l=0,\quad k=1,\ldots, n<m,\quad \mathrm{rang}\,a(x)=n.\end{equation}
{\bf Here and below use the Einstein summation convention.} We also use the notation $c,c_1,c_2\ldots$ for inessential positive constants.

The function $T$ is the kinetic energy of the system; it is a positive definite quadric in the variables $\dot x$.  The matrix $G$ is positive definite and determines a Riemann metric in $M$. The function $V$ is a potential. 

We watch the motion of this system on the time interval $ I_\tau=[0,\tau].$

In the classical situation all these functions are smooth in $M$ and the dynamics of the system is described by the Lagrange-d'Alembert principle 
\begin{equation}\label{v5t5}\Big(\frac{d}{dt}\frac{\partial L}{\partial \dot x^j}-\frac{\partial L}{\partial x^j}\Big)\psi^j=0. \end{equation}

By definition  the function $x(t)$ is a motion of the system iff it satisfies  (\ref{vt55v}) and for any functions $\{\psi^j(t)\}$ such that $a^k_l(x(t))\psi^l(t)=0$ it satisfies (\ref{v5t5}).

To present system (\ref{vt55v})-(\ref{v5t5}) in the resolved with respect to the highest derivatives form one must rewrite this system using Lagrange multipliers:
\begin{equation}\label{ervb45yb}
\frac{d}{dt}\frac{\partial L}{\partial \dot x}-\frac{\partial L}{\partial x}=\Lambda(x,\dot x) a(x),\quad \Lambda(x,\dot x)=(\lambda_{1},\ldots,\lambda_{n})(x,\dot x).\end{equation}
To express the Lagrange multipliers $\lambda$ one should take time derivative from the both sides of (\ref{vt55v}) and substitute there $\ddot x$ from 
(\ref{ervb45yb}).

After these transformations one obtains 
\begin{equation}\label{evt54vgg}\Lambda^T(x,\dot x)=\Big(a(x)G^{-1}(x)a^T(x)\Big)^{-1} w\Big(\frac{\partial a}{\partial x},\frac{\partial G}{\partial x},a,G,x,\dot x\Big).\end{equation}
The function $w$ is smooth in $x\in M$, and in the other arguments $w$ is smooth in the whole space. Correspondingly,  (\ref{vt55v}) is the equation of an invariant manifold to system (\ref{ervb45yb}).

Assume that \begin{equation}\label{tbt}G,a\in H^{1,\infty}(M).\end{equation} This particularly implies 
$G(x),a(x)\in C(M).$

This situation, for example, takes place when  the Chaplygin sleigh  \cite{nem-fuf} moves on a very irregular  surface, say $z=f(x,y)$ and the function $f$ is constructed as follows. Let $\varphi(x)$
be a smooth function with compact support and such that $\varphi(x)=1$ for $x\in(-\delta,\delta).$ Then, we let
$$f(x,y)=\sum_{k=1}^\infty\frac{1}{2^k}\varphi(x-x_k)(x-x_k)^4\cos\Big(\frac{1}{x-x_k}\Big)$$
and the sequence $\{x_k\}_{k\in\mathbb{N}}$ runs over all rationals $\mathbb{Q}$.

In such a case the Lagrange multipliers (\ref{evt54vgg}) cannot be defined correctly because the function $$\frac{\partial a}{\partial x}$$ belongs just to $L^\infty(M)$ and even it is  not clear what   the expression
$$\frac{\partial a}{\partial x}(x(t))$$ means.
Therefore, equation (\ref{ervb45yb}) also becomes  impossible.

In this article we propose a definition of weak solutions to the Lagrange-d'Alembert equations. This definition allows us to overcome the described problem and prove an existence theorem for the weak solutions. Nevertheless the question on the uniqueness remains open.

For general dynamical systems $\dot x=f(t,x)$ with  non-Lipschitz $f$ there are a lot of works devoted to investigating 
different types of  uniqueness conditions in case.
As far as the author knows this activity has been started from Kamke \cite{kamke} 
and Levy \cite{levy}. Their results have been generalized in different directions. 
See for example \cite{ramankutty}, \cite{bownds1}
 and references therein.

The case when $f$  belongs to Sobolev spaces (at least $H^{1,1}$)
has been studied in \cite{DiP-Lions} in connection with the Navier-Stokes equation.
In that article the equations which have good invariant measure are mainly considered. The existence and uniqueness theorems for the flow are given in terms of the corresponding transport equation.

In this article we consider individual solutions to the Cauchy problem for general non-holonomic systems; such systems need not necessarily possess a good invariant measure.

Collisions in holonomic Lagrangian systems have been considered in \cite{KozlovTreschev}. Collisions provide  a source of another type of singularities and  generalized solutions in dynamics.

\section{Main Theorem}

Let  $\|\cdot\|$ stand for the $l_2$-norm in $\mathbb{R}^m$.
We introduce the following subspace of the Sobolev space $H^1(I_\tau)$
$$H_0^1(I_\tau)=\{u\in H^1(I_\tau)\mid u(0)=0\}.$$

In the sequel  by $c, c_1,c_2\ldots$ we denote  positive constants.

Give a precise description of our functions: $V\in C^2(M)$ and the functions $g_{ij}$  are such that for almost all $x\in M$ the conditions  $$g_{ij}(x)=g_{ji}(x) ,\quad c_1\|\xi\|^2\le g_{ij}(x)\xi^i\xi^j\le c_2\|\xi\|^2,\quad \xi\in\mathbb{R}^m$$ hold.

 We also suppose that for some constant $h$ a domain $$D_h=\{x\in M\mid V(x)<h\}$$ is non void and $\overline D_h\subset M$.

A non degeneracy condition is also applied:
\begin{equation}\label{dry5}A(x)=\big(a^k_l(x)\big)_{k,l=1,\ldots,n},\quad \det A(x)\ne 0,\quad x\in M.\end{equation}

We introduce the energy of the system $H(x,\dot x)=T(x,\dot x)+V(x).$
 
Integrating (\ref{v5t5}) by parts one obtains   the Lagrange-d'Alembert principle  in the integral form \cite{bloch}. This justifies the following definition.
\begin{df}\label{tvbgtr}We shall say that a function $x(t)\in H^1(I_\tau)$ is a weak solution to the system of Lagrange-d'Alembert equations and the equations of constraint iff the equation
\begin{align}\int_{I_\tau}\Big(\frac{\partial L}{\partial x }(x(t),\dot x(t))\psi(t)&+\frac{\partial L}{\partial \dot x}(x(t),\dot x(t))\dot\psi(t)\Big)\,dt\nonumber\\&-\frac{\partial L}{\partial\dot x}(x(\tau),\dot x(\tau))\psi(\tau)=0\label{c34rt5}\end{align}
holds for any $\psi(t)=(\psi^1,\ldots,\psi^m)^T(t)\in H^1_0(I_\tau)$ that  satisfy 
\begin{equation}\label{5bgy45yff}a(x(t)) \psi(t)=0,\end{equation}
and equation (\ref{vt55v})  holds for almost all $t\in I_\tau$ that is, $a(x(t))\dot x(t)=0.$
\end{df}
Observe that due to compact embedding $H^1(I_\tau)\subset C(I_\tau)$ this 
definition implies $x(t)\in C(I_\tau)$.

\begin{theorem}\label{5bvy46y}

For any positive constant $\tau$ and for any initial conditions $x_0,v$ such that
$$a(x_0)v=0$$ and
\begin{equation}\label{qqqqqqqqqqqq}H(x_0,v)=h'<h\end{equation}
there exists a weak solution $x(t),\quad x(0)=x_0,\quad \dot x(0)=v$ to the Lagrange-d'Alembert equations and the equations of constraint (\ref{vt55v}).

Moreover $H(x(t),\dot x(t))=h' \forall \quad t\in I_\tau$ and $x\in C^{1,\alpha}(I_\tau)$ for any $\alpha\in (0,1)$.\end{theorem}

\begin{theorem}\label{bygh6ye|}
Let $x(t)$ be a solution to the system of Lagrange-d'Alembert equations in the sense of Definition  \ref{tvbgtr}.

Then there exists a function $\gamma(t)=(\gamma_1,\ldots,\gamma_n)(t)\in L^2(I_\tau)$ such that the equation
\begin{align}\int_{I_\tau}&\Big(\frac{\partial L}{\partial x }(x(t),\dot x(t))\psi(t)+\frac{\partial L}{\partial \dot x}(x(t),\dot x(t))\dot\psi(t)\Big)\,dt\nonumber\\&-\frac{\partial L}{\partial\dot x}(x(\tau),\dot x(\tau))\psi(\tau)=\int_{I_\tau}\gamma(t)a(x(t))\psi(t)\,dt\label{c3rr4rt5}\end{align}

holds for any $\psi\in H^1_0(I_\tau)$.
\end{theorem}

\section{Proof of Theorem \ref{5bvy46y}}\label{etvb5t}
Introduce  matrices
\begin{align}Q(x)&=\big(a^k_l(x)\big),\quad l=n+1,\ldots,m,\quad k=1,\dots,n,\nonumber\\ B(x)&=-A^{-1}(x)Q(x).\nonumber\end{align}

\begin{prop}\label{5evbgy4b}
Suppose that for some $x_0\in M$ and $v\in\mathbb{R}^m$ one has $a(x_0)v=0$.
Then there is a sequence $G_i(x),a_{i}(x)\in C^\infty(M)$ such that
\begin{equation}\label{gtbr6yy6}\|G_i-G\|_{L^\infty(M)},\quad\|a_{i}-a\|_{_{L^\infty(M)}}\to 0\quad \mbox{as}\quad i\to\infty,\end{equation}
and $a_i(x_0)v=0,$
$$\Big\|\frac{\partial G_{i}}{\partial x}\Big\|_{L^\infty(M)},\quad\Big\|\frac{\partial a_{i}}{\partial x}\Big\|_{L^\infty(M)}\le c.$$
\end{prop}{\it Proof.}
First let us recall a standard fact.

There is a sequence $a^*_{i}(x)\in C^\infty(M)$ such that
$$\|a^*_{i}-a\|_{_{L^\infty(M)}}\to 0\quad \mbox{as}\quad i\to\infty,$$
and
$$\Big\|\frac{\partial a^*_{i}}{\partial x}\Big\|_{L^\infty(M)}\le c.$$
The constant $c$ does not depend on $i$.
This  follows from real analysis and formula (\ref{tbt}).

Thus  if we find a sequence $\{b_i\}$ such that
$$\|b_i\|\to 0,\quad b_iv=-a^*_i(x_0)v$$ 
and put $a_i(x)=a_i^*(x)+b_i$ then the Proposition is proved.

Let $v=(v^1,\ldots,v^m)^T$ and  $v^1\ne 0$ furthermore  $$b_i=(b_{ij}^r),\quad a^*_i(x_0)=(w_{il}^k)$$
and observe that $w_{il}^kv^l\to 0,\quad i\to\infty$.

It remains to take $b_{ij}^r=0,\quad j>1$ and
$$ b_{i1}^k=-\frac{w_{il}^kv^l}{v^1}.$$

The Proposition is proved.

Let us approximate our initial problem by the smooth problems:

\begin{align}
\frac{d}{dt}\frac{\partial L_i}{\partial \dot x}-\frac{\partial L_i}{\partial x}&=\Lambda_i(x,\dot x) a_i(x),\quad \Lambda_i(x,\dot x)=(\lambda_{i1},\ldots,\lambda_{in})(x,\dot x),\label{5vb5dfg345}\\
a_i(x)\dot x&=0,\quad L_i=\frac{1}{2}\dot x^TG_i(x)\dot x-V(x). \label{5vb534sdf}
\end{align}
To express the Lagrange multipliers $\lambda$ one should take time derivative from the both sides of (\ref{5vb534sdf}) and substitute there $\ddot x$ from (\ref {5vb5dfg345}).

After these transformations one obtains 
\begin{equation}\label{r6buy5ugg}\Lambda^T_i(x,\dot x)=\Big(a_i(x)G_i^{-1}(x)a_i^T(x)\Big)^{-1} w\Big(\frac{\partial a_i}{\partial x},\frac{\partial G_i}{\partial x},a_i,G_i,x,\dot x\Big).\end{equation}
The function $w$ is smooth in $x\in M$, and in the other arguments $w$ is smooth in the whole space. 

Meanwhile system (\ref{5vb5dfg345}) with formula (\ref{r6buy5ugg}) takes the form
\begin{equation}\label{d5ybv56y}\ddot x=\phi_i(x,\dot x),\end{equation}
where $\phi_i\in C^\infty( M\times \mathbb{R}^m).$ 
Equation (\ref{5vb534sdf}) determines an invariant manifold to system (\ref{d5ybv56y}). 

Recall that  Proposition \ref{5evbgy4b} implies $a_i(x_0)v=0$.

The keypoint of our argument is as follows: system (\ref{5vb5dfg345})-(\ref{5vb534sdf}) possesses the energy integral $H$, thus the function $H$ is also the first integral for system (\ref{d5ybv56y}).

Summarize   the above argument  as a lemma.
\begin{lemma}\label{etgb45t}
For the constant $\tau>0$ and for the initial conditions $$x_i(0)=x_0,\quad \dot x_i(0)=v$$ 
system (\ref{d5ybv56y}) has a solution $x_i(t)\in C^2(I_\tau)$ such that 
\begin{equation}\label{dr5yb4}H(x_i(t),\dot x_i(t))=h'.\end{equation}
and $a_i(x_i(t))\dot x_i(t)=0,\quad \forall t\in I_\tau$.
\end{lemma}

\begin{cor}\label{e5vb6} The sequence  $\{x_i(t)\}$ contains a subsequence that is convergent in $C^{1,\alpha}(I_\tau)$. 

For this subsequence we use the same notation, that is $$\|x_i-x\|_{C^{1,\alpha}(I_\tau)}\to 0.$$\end{cor} 
Indeed, combining  Proposition \ref{5evbgy4b} and formulas (\ref{dr5yb4}), (\ref{r6buy5ugg}), (\ref{5vb5dfg345})  one has
\begin{equation}\label{5by6yb456ffr}\|\phi_i(x_i(s,v_i),\dot x_i(s,v_i))\|\le K.\end{equation}
The constant $K$ is independent of $t\in (I_\tau)$ and $i$.
Then from  Lemma \ref{etgb45t} and formulas (\ref{d5ybv56y}), (\ref{5by6yb456ffr}) one concludes that the sequence $\{\ddot x_i(t)\}$ is uniformly bounded in $(I_\tau)$. 

\begin{lemma}\label{ev5by4yb} The function $x(t)$ from  Corollary \ref{e5vb6} satisfies (\ref{vt55v}).\end{lemma}
{\it Proof.} Follows directly from Proposition \ref{5evbgy4b} and Corollary \ref{e5vb6}.

Introduce matrices $$A_i(x)=\big(a_{il}^k(x)\big)_{k,l=1,\ldots, n}.$$
\begin{lemma}\label{drtgyff}
The following estimate holds
$$\sup_{i}
\Big\|\frac{\partial A_i^{-1}(x)}{\partial x}\Big\|_{L^\infty(D_h)}<c_3.$$\end{lemma}
{\it Proof.}
Differentiate the $A_i^{-1}A_i=I$ by $x^s$:
$$\frac{\partial A_i^{-1}}{\partial x^s}=-A^{-1}_i\frac{\partial A_i}{\partial x^s}A_i^{-1}.$$
Now the assertion follows from Proposition \ref{5evbgy4b} and assumption (\ref{dry5}):
$$\Big\|\frac{\partial A_i^{-1}}{\partial x^s}\Big\|_{L^\infty(D_h)}\le \Big\|A^{-1}_i\Big\|_{L^\infty(D_h)}^2\cdot\Big\|\frac{\partial A_i}{\partial x^s}\Big\|_{L^\infty(D_h)}.$$

The Lemma is proved.

Observe also that since $A_i$ is uniformly closed to $A$ and $\det A(x)\ge c_3>0,\quad x\in D_{h}$ one obtains $\|A_i^{-1}(x)\|_{L^\infty(D_{h})}\le c_4$ for some $c_4>0$ if only $i$ is sufficiently large.

Consider spaces 
$$ E_i=\{\psi\in H^1_0(I_\tau)\mid a_i(x_i(t))\psi(t)=0\}.$$

\begin{lemma}\label{rybr6y56yfdfff}
For any $\psi\in E$ there exist a sequence $\{\psi_i\},\quad \psi_i\in E_i$ such that $\psi_i\to \psi$ weakly in $H^1_0(I_\tau)$ and strongly in $C(I_\tau)$.\end{lemma}

{\it Proof.}
Introduce  matrices
$$ Q_i(x)=\big(a_{il}^k(x)\big),\quad l=n+1,\ldots,m,\quad k=1,\dots,n.$$

Fix an arbitrary function  $$ \hat \psi=( \hat\psi^{n+1},\ldots, \hat\psi^m)^T\in H^1_0(I_\tau).$$

Consider a sequence $$\tilde \psi_i(t)=-A_i^{-1}(x_i(t))Q_i(x_i(t))\hat \psi(t).$$ This sequence is bounded in $C[0,1]$.
By Lemma \ref{drtgyff} the sequence 
\begin{align}\frac{d}{dt}\tilde \psi_i(t)&=-\frac{\partial A_i^{-1}(x_i(t))}{\partial x^l}\dot x^l(t)Q_i(x_i(t))\hat\psi(t)\nonumber\\&- A_i^{-1}(x_i(t))\dot x^l(t)\frac{\partial Q_i(x_i(t))}{\partial x^l}\dot x^l(t)\hat\psi(t)\nonumber\\
&-A_i^{-1}(x_i(t))Q_i(x_i(t))\frac{d}{dt}\hat\psi(t)\nonumber\end{align}
 is bounded in $L^2(I_\tau)$.

So, using the same notation for subsequences we have  $\tilde \psi_i(t)\to \tilde\psi(t)$ weakly in $H^1(I_\tau)$. Convergence in $C(I_\tau)$ follows from compact embedding $H^1(I_\tau)\subset C(I_\tau)$.

We want to pass to the limit as $i\to \infty$ in the equality
$$A_i(x_i(t))\tilde \psi_i(t)+Q_i(x_i(t))\hat\psi(t)=0.$$ Since $H^1(I_\tau)$ is compactly embedded in $C[0,1]$ the sequence $\tilde\psi_i$ converges to $\tilde \psi(t)$ in $C(I_\tau)$. Thus we have
$$A(x(t))\tilde\psi(t)+Q(x(t))\hat\psi(t)=0.$$ 

Thus the sequence we are looking for is
$$\psi_i=(\tilde \psi^1_i,\ldots,\tilde\psi^n_i,\hat\psi^{n+1},\ldots,\hat\psi^m)^T.$$

The Lemma is proved.

Let us observe another evident fact.
\begin{lemma}\label{e5bg646}
Suppose that   a sequence $\{u_i\}\in L^2(I_\tau)$ converges weakly to $u\in L^2(I_\tau)$.
We also have a sequence of  functions $\{f_i\}\subset C(I_\tau)$. This sequence converges uniformly to $f\in C(I_\tau)$.

Then $$( f_i, u_i)_{L^2(I_\tau)} \to ( f, u)_{L^2(I_\tau)}.$$ 
\end{lemma} 
{\it Proof.} Indeed,  one has
$$( f_i, u_i)_{L^2(I_\tau)}=(f_i-f,u_i)_{L^2(I_\tau)}+(f,u_i)_{L^2(I_\tau)}$$
and since the sequence $\{u_i\}$ is bounded in $L^2(I_\tau)$ \cite{yosida} it follows that $$|(f_i-f,u_i)_{L^2(I_\tau)}|\le\|f_i-f\|_{C(I_\tau)}\|u_i\|_{L^2(I_\tau)}\to 0.$$

The Lemma is proved.

\begin{lemma}\label{d5bg5y45ffff}
Take any $\psi\in E$ and choose the sequence $\psi_i$ in accordance with Lemma  \ref{rybr6y56yfdfff}.
Then 
\begin{align}
\int_{I_\tau}\frac{\partial L}{\partial \dot x}\big(x_i(t),\dot x_i(t)\big)\dot\psi_i(t)\,dt&\to \int_{I_\tau}\frac{\partial L}{\partial \dot x}\big(x(t),\dot x(t)\big)\dot\psi(t)\,dt\label{srvb456}\\
\int_{I_\tau}\frac{\partial L}{\partial x }\big(x_i(t),\dot x_i(t)\big)\psi_i(t)dt&\to \int_{I_\tau}\frac{\partial L}{\partial x }\big(x(t),\dot x(t)\big)\psi(t)\,dt.\label{rbdd}\end{align}
\end{lemma}
{\it Proof.} Limit (\ref {rbdd}) is trivial. Let us prove formula (\ref{srvb456}).
Since \begin{align}\int_{I_\tau}\frac{\partial L}{\partial \dot x}\big(x(t),\dot x(t)\big)\dot\psi(t)\,dt&=\big(\dot x^T(\cdot)G(x(\cdot)),\dot\psi(\cdot)\big)_{L^2(I_\tau)},\nonumber\\
\int_{I_\tau}\frac{\partial L}{\partial \dot x}\big(x_i(t),\dot x_i(t)\big)\dot\psi_i(t)\,dt&=\big(\dot x_i^T(\cdot)G(x_i(\cdot)),\dot\psi_i(\cdot)\big)_{L^2(I_\tau)}\nonumber\end{align}
the  assertion of the Lemma follows from Lemma \ref{e5bg646}.

The Lemma is proved.

It remains to observe that the existence in  Theorem \ref{5bvy46y} follows directly from Lemma \ref{d5bg5y45ffff}.

The Theorem is proved.

\section{Proof of the Theorem \ref{bygh6ye|}} Introduce the following spaces
\begin{align}X&=\{\psi=(\psi^1,\ldots,\psi^m)^T\mid \psi^k\in H^1_0(I_\tau)\},\quad \|\cdot\|_X=\|\cdot\|_{L^2(I_\tau)},\nonumber\\
Y&=\{\varphi=(\varphi^1,\ldots,\varphi^n)^T\mid \varphi^k\in L^2(I_\tau)\},\quad\|\cdot\|_Y=\|\cdot\|_{L^2(I_\tau)}.\nonumber\end{align}
Note that the  space $X$ is not a Banach space.

Let $S:X\to Y$ stand for the operator $\psi(t)\mapsto a(x(t))\psi(t)$.
 $F:X\to\mathbb{R}$ stands for the linear functional
\begin{align}\psi\mapsto\int_{I_\tau}&\Big(\frac{\partial L}{\partial x }
(x(t),\dot x(t))\psi(t)+\frac{\partial L}{\partial \dot x}(x(t),\dot x(t))\dot\psi(t)\Big)\,dt\nonumber\\&-\frac{\partial L}{\partial\dot x}(x(\tau),\dot x(\tau))\psi(\tau).\nonumber\end{align}
We know that $\ker S\subseteq\ker F$, let us check inequality (\ref{dbyvbv}).

For a vector $y=(y^1,\ldots,y^m)^T\in\mathbb{R}^m$ introduce operations
$$\tilde y=(y^1,\ldots, y^n)^T,\quad \hat y=(y^{n+1},\ldots, y^m)^T.$$

Let $\psi\in X$ then put 
$$\tilde \psi_o(t)=B(x(\tau))\hat\psi(t),\quad \psi_o(t)=(\tilde \psi_o^T(t), \hat \psi(t)^T)^T,\quad \psi_\dag(t)=\psi(t)-\psi_o(t)$$
so as $$\psi_o\in\ker S,\quad \psi=\psi_o+\psi_\dag,\quad \hat{\psi_\dag}=0.$$

So we have $$\|S(x(\cdot))\psi(\cdot)\|_{L^2(I_\tau)}^2=\|A(x(\cdot))\tilde{\psi_\dag}(\cdot)\|^2_{L^2(I_\tau)}\ge c_8\|\psi_\dag(\cdot)\|^2_{L^2(I_\tau)}$$ for some $c_8>0$
and finally one yields  $$\|\psi_\dag(\cdot)\|^2_{L^2(I_\tau)}\ge \inf_{\nu\in\ker S}\|\psi(\cdot)+\nu(\cdot)\|^2_{L^2(I_\tau)}.$$

By Lemma \ref{dtbvr6tby} we have a bounded functional $$\Gamma:S(X)\to\mathbb{R},\quad F=\Gamma S,\quad \|\Gamma\|\le\frac{1}{c_8}\|F\|.$$ Using the Hahn-Banach theorem, we extend this functional to a bounded functional  $\Gamma_1:Y\to \mathbb{R}$.

By the Riesz Representation Theorem one can find a function $\gamma(\tau)=(\gamma_1,\ldots,\gamma_n)(\tau),\quad \gamma_k\in L^2(I_\tau)$ such that $$\Gamma_1\varphi=(\gamma,\varphi)_{L^2(I_\tau)},\quad \|\gamma\|_{L^2(I_\tau)} \le\frac{1}{c_8}\|F\|.$$

The Theorem is proved.

\section{A Lemma from Functional Analysis}The following lemma is well known. We bring its proof just for completeness of exposition.

Let $X,Y,Z$ be normed spaces and linear operators $$F:X\to Z,\quad S:X\to Y$$ be  bounded.
\begin{lemma}\label{dtbvr6tby}
Suppose that $\ker S\subseteq\ker F$ and \begin{equation}\label{dbyvbv}\inf_{u\in \ker S}\|z+u\|_X\le C\|Sz\|_Y, \end{equation} for some $C>0$.

Then there is a bounded operator $\Gamma:S(X)\to Z$ such that 
$$F=\Gamma S,\quad \|\Gamma\|\le C\|F\|.$$\end{lemma}
{\it Proof.} 
Let $$\pi_S:X\to V=X/\ker S,\quad \pi_F:X\to U=X/\ker F,\quad \pi:V\to U$$
be natural projections.

The spaces $U,V$ are normed spaces with norms $$\|u\|_U=\inf_{w\in\ker F}\|[u]+w\|_X,\quad \|v\|_V=\inf_{w\in\ker S}\|[v]+w\|_X$$ where $[u]\in X$ is the element that generates corresponding class $u$ that is $u=\ker F+[u]$.

From \cite{Rob} we know that $F=F_1\pi_F,\quad S=S_1\pi_S$ and the bounded operators $F_1:U\to F(X),\quad S_1:V\to S(X)$ are one-to-one. 

Hence we have $\Gamma=F_1\pi S_1^{-1}$. By formula (\ref{dbyvbv}) the operator
$S_1^{-1}$ is bounded.

The Lemma is proved.

\subsection*{Acknowledgement}

%I wish to thank Professor D.V. Treschev for  drawing  my attention to this %topic and for useful discussions.

Partially supported by grants
 RFBR 12-01-00441,  Science Sch-2964.2014.1.

\end{document}